\documentclass[10pt,aps,prb,twocolumn,amsmath,amssymb,amsfonts,longbibliography,floatfix,
    superscriptaddress,nobibnotes]{revtex4-2} 
\usepackage[utf8]{inputenc}
\usepackage[T1]{fontenc} 
\usepackage{ebgaramond}
\usepackage{braket}
\usepackage{graphicx} 
\usepackage[version=4]{mhchem}
\usepackage{tikz}
\usepackage{times}
\usepackage{soul}
\usepackage{url}
\usepackage{mathptmx}
\usepackage{gensymb}
\usepackage{todonotes}
\usepackage{verbatim}
\usepackage{xcolor}
\usepackage{hyperref}
\hypersetup{
  hypertexnames=false,
  colorlinks=true,
  citecolor=blue,
  linkcolor=blue,
  urlcolor=blue
}

\definecolor{mygreen}{rgb}{0.0, 0.6, 0.0}

\DeclareMathAlphabet\mathbfcal{OMS}{cmsy}{b}{n}

\newcommand{\opt}[1]{%
\begin{tikzpicture}[baseline=(current bounding box.center)]
\node[fill=gray!20,inner sep=4.0pt,rounded corners=1pt] {#1};
\end{tikzpicture}%
}

\graphicspath{{fig/}{./fig/}{.}}

\begin{document}

\title{Controlling spin-$\frac 12$ antiferromagnetic interaction strength in nanographene dimers}

\author{Robiatul Adawia}
\email[e-mail: ]{adawiarobiatul@gmail.com}
\affiliation{Institute of Physics, Faculty of Physics, Astronomy and Informatics, Nicolaus Copernicus University in Toruń, Grudziadzka 5, 87-100 Toru\'n, Poland.}

\author{Pawe\l~Tecmer}
\affiliation{Institute of Physics, Faculty of Physics, Astronomy and Informatics, Nicolaus Copernicus University in Toruń, Grudziadzka 5, 87-100 Toru\'n, Poland.}

\author{Pawe\l~Potasz}
\affiliation{Institute of Physics, Faculty of Physics, Astronomy and Informatics, Nicolaus Copernicus University in Toruń, Grudziadzka 5, 87-100 Toru\'n, Poland.}

\date{\today}

\begin{abstract}
We demonstrate that the effective spin-exchange coupling $J$ in open-shell nanographene dimers can be precisely tuned via tip-induced dehydrogenation of selected carbon atoms.
Using the double ionization potential equation-of-motion coupled-cluster singles and doubles (DIP-EOM-CCSD) method, we accurately compute the singlet–triplet gaps, which correspond directly to the exchange coupling $J$. 
We show that the position of the dehydrogenated (or hydrogen-passivated) site in triangulene dimers strongly modulates the singlet–triplet splitting, allowing $J$ to be tuned over a wide range — from a few meV to several tens of meV.
This strategy provides a simple yet powerful route for designing tailored spin models with alternating or spatially patterned spin-exchange couplings.
\end{abstract}

\maketitle


\section{Introduction}

Open-shell nanographenes have emerged as a promising platform to realize carbon-based magnetism and quantum simulators of spin models \cite{yazyev2008magnetic, yazyev2010emergence, han2014graphene,ortiz2019fernandez,mishra2021large, de2022carbon}. Contrary to d/f electron magnetism in transition metal elements, the low atomic
mass of carbons ensures negligible magnetic anisotropy, together with weak spin-orbit interaction translates to long coherence times in potential future quantum devices \cite{min2006intrinsic,yazyev2008hyperfine,yazyev2008magnetic}. While for years it was challenging to synthesize nanographenes hosting $\pi$-magnetism due to high reactivity of molecules with unpaired electrons,  recent advancements in on-surface synthesis have enabled the fabrication of a number of magnetic nanographenes on a Au(111)
surface \cite{ruffieux2016surface,mishra2020collective,mishra2021large,song2021surface, mishra2020topological}. Among many finite size structures, these includes Clar’s goblet \cite{mishra2020topological}, [n]-triangulenes \cite{pavlivcek2017synthesis,su2019atomically, mishra2019synthesis,mishra2021synthesis,turco2023observation}, aza-[n]triangulenes \cite{wang2022aza,lawrence2023topological,rodrigues2025probing}, a triangulene ring \cite{su2020surface},
[n]-rhombenes \cite{mishra2021large,biswas2023steering}, nanographene trimers \cite{cheng2022surface,du2023orbital} and opened the door for realization of quantum simulators of various spin chain models \cite{mishra2021observation,zhao2024tunable,zhao2025spin,fu2025building,su2025fabrication}. 

Although isolated [n]-triangulenes reveal a finite magnetic moment \cite{graphenebook,Ezawa2007,fernandez2007magnetism,Wang2008NL,GuccluPRL2009,Potasz2010ZeroEnergy, potasz2012electronic}, coupled structures can form spin molecules with either large or minimal total spin, depending on the inter-triangulene connection form \cite{mishra2020collective}. 
The ground-state spin of nanographenes is generally described by the Ovchinnikov-Lieb rule \cite{ovchinnikov1978multiplicity,lieb1989two}, which relates it to an imbalance betweeen two sublattices forming a graphene lattice, $S=|N_A-N_B|$, where $N_A$ and $N_B$ is the number of lattice sites of each sublattice. However, a distinction between non-magnetic and open-shell
singlet ground state systems, both corresponding to minimal total spin $S=0$ and $N_A=N_B$, requires a determination of a number of singly occupied
molecular orbitals. This is resolved by the topology of nanograpehene, described by the nullity, number of nonbonding $\pi$ orbitals \cite{alma991000470629706691,fajtlowicz2005maximum,Wang2008NL}. The nullity of the graph determines the number of zero-energy modes within the nearest neighbor tight-binding model. The simplest examples are singlet diradicals, molecules with two unpaired electrons that are not forming a conventional bond, giving it unusual bonding and high reactivity. In other words, for systems with total spin $S=0$ and nonzero nullity, unpaired $\pi$ electrons filling these zero-energy states lead to local asymmetries in spin densities in different parts of the structure and the effective magnetic interaction arises. 

Designing antiferromagnetic spin coupling is of particular interest due to highly entangled nature of low total spin ground states and the possible emergence of exotic quantum phases \cite{pitaevskii1991uncertainty,anderson1973resonating,balents2010spin,savary2017quantum,takagi2019concept,chamorro2020chemistry}. Nanographenes as building blocks of such systems are especially attractive in this context, as magnetic exchange interactions can also be tuned through molecular design or chemical functionalization  \cite{mishra2020topological,wang2022aza,krane2023exchange,catarina2024conformational,zhao2024tailoring,derradji2025functionalization}. Within these methods, tip-induced dehydrogenation has enabled site-specific manipulation of magnetic moments in Clar’s goblet by creating carbon vacancies \cite{zhao2024tailoring}, thereby redistributing the spin density of unpaired electrons. This is related to created Au—C bond, as the effect of strong hybridization of carbon orbitals with the Au substrate states. Following this idea, we investigate how the magnetic exchange interaction can be modified in [3]-triangulene dimers by selecting a tip-induced dehydrogenated carbon site.

A simplified approach to estimating the strength of the exchange interaction consists in including only singly occupied molecular orbitals in the complete active space (CAS) calculation. 
However, one of us in Ref. \onlinecite{saleem2024superexchange} has demonstrated that the coupling arises from a superexchange mechanism, and that, additionally, molecular orbitals localized at the junction between the structures must also be included.
As a result, obtaining reliable results within CAS calculations may require the inclusion of a large number of molecular orbitals in the active space; at the same time, this space must remain limited due to the high computational cost of the method. This becomes particularly problematic when many molecular orbitals are close in energy and none can be safely neglected.
Similarly, the low-energy many-body states exhibit a multiconfigurational character and cannot be accurately described by density functional theory (DFT).
These considerations motivate us to adopt an alternative approach that, on the one hand, accounts for static electron correlation (i.e., a multiple Slater determinant configuration method) and, on the other hand, does not restrict the set of included single-particle molecular orbitals. 
To this end, we employ the double ionization potential equation-of-motion coupled-cluster (DIP-EOM-CCSD) formalism to accurately compute singlet–triplet gaps in open-shell nanographene dimers.
Specifically, we demonstrate that the magnetic exchange interaction strength in these spin-$\frac{1}{2}$ dimers can be tuned from approximately 10 meV to nearly 90 meV by the strategic choice of dehydrogenated carbon sites.
The reliability of our approach is further confirmed by the excellent agreement between the computed singlet–triplet gaps for olympicenes and experimental measurements.
\section{Description of open-shell states with DIP-EOM-CCSD}\label{sec:dipeomccsd}
Our starting point is the coupled cluster singles and doubles ansatz~\cite{cizek-jcp-1966, cizek-paldus-1971, bartlett-review-arpc-1981, bartlett-review-rmp-2007, helgaker-book-2000, shavitt-bartlett-book-2009} on top of Hartree--Fock (HF) reference wave function $|\Phi_0\rangle$ defined as
\begin{equation}
    |{\rm CCSD}\rangle  = e^{\hat{T}^{\rm CCSD}}|\Phi_0\rangle = e^{\hat{T}_{1} +\hat{T_2}}|\Phi_0\rangle, 
\end{equation}
where $\hat{T}_{1}$ and $\hat{T}_{2}$ are the cluster operators containing single and double electron excitations. 
Here, we restrict ourself to the spin-free case, where all excitation operators are written in terms of singlet excitation operators $\hat E_a^i$,
\begin{equation}
\label{eq:singlet_excitation_op}
    \hat E_a^i = \hat a_{a}^\dagger \hat a_{i} + \hat a_{\bar a}^\dagger \hat a_{\bar i}.
\end{equation}
For spin-free double excitations, the $\hat T_2$ cluster operator takes on the form
\begin{equation}
    \hat T_2 = \frac{1}{2} \sum_{i j a b} t_{ij}^{ab} \, \hat E_{a}^{i} \hat E_{b}^{j},
    \label{eq:ccd_spinadapted}
\end{equation}
while single excitations reduce to
\begin{equation}
    \hat T_1 = \sum_{i a} t_{i}^{a} \, \hat E_{a}^{i}.
    \label{eq:ccs_spinadapted}
\end{equation}

We can utilize the equation of motion (EOM) formalism
on top of the CCSD reference, 
\begin{equation}\label{eq:eom-pccd-fpcc}
{[\hat{H}_N,\hat R ]} \ket{\textrm{CCSD}} =  \omega \hat R \ket{\textrm{CCSD}},
\end{equation}
to obtain electronically excited,~\cite{stanton-bartlett-eom-jcp-1993, bartlett-eom-cc-wires-2012, eom-pccd, eom-pccd-erratum, eom-pccd-lccsd}
spin-flip,~\cite{spin-flip-eom-casanova-pccp-2020} 
attached,~\cite{nooijen-ea-eom-jcp-1995, musial2014equation, ip-pccd, dea-ea-pccd-fpcc-jpca-2024} 
ionized,~\cite{musial2003equation, ip-pccd, ip-tailoredpccd-jctc-2024}
doubly attached, and doubly ionized states.~\cite{DIP-EOM1, DIP-EOM2, dea-ea-pccd-fpcc-jpca-2024, dip-eomfpccsd-jctc-2025} 
In equation \eqref{eq:eom-pccd-fpcc}, $\hat{H}_N$ represents the Hamiltonian in its normal-product form,
$ \omega = \Delta E_k - \Delta E_0$ denotes the difference between the ground- and the $k$-th (excited, spin-flip, (doubly) attached, or (doubly) ionized) state associated with a specific form of the linear $\hat{R}$ operator 
\begin{equation}
\ket{\Psi_k} = \hat R(k)\ket{\textrm{CCSD}}.
\end{equation}

\begin{figure}[h]
\centering
\includegraphics[width=0.8\columnwidth]{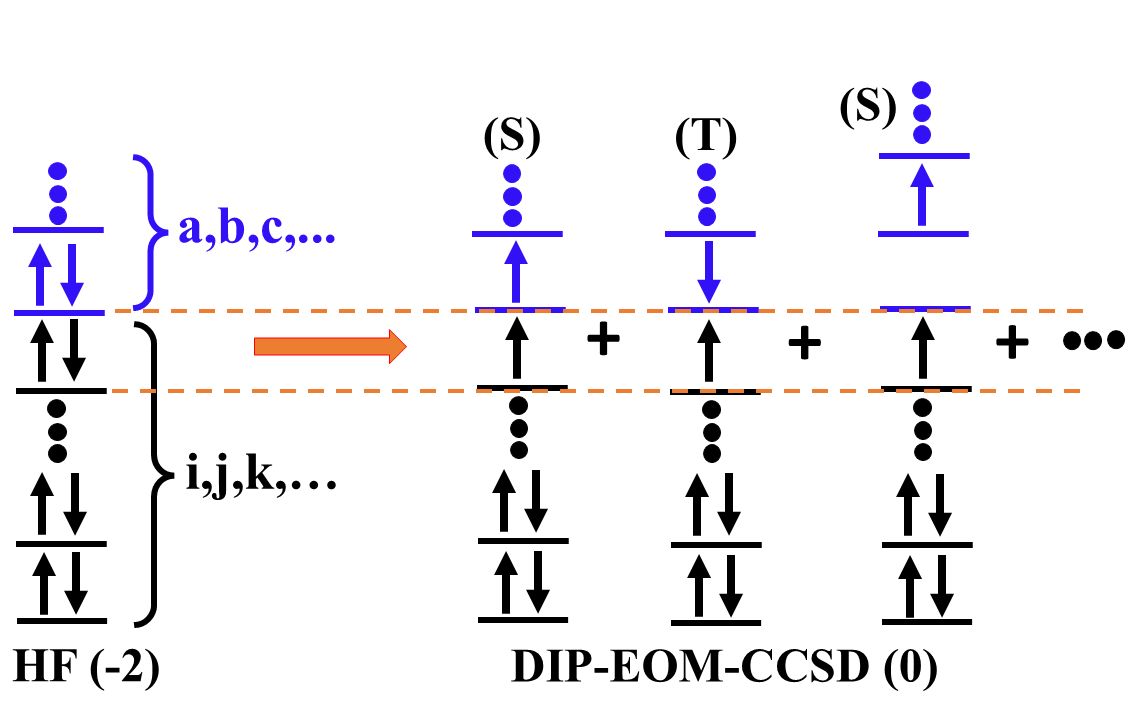}
\caption{Schematic representation of the DIP-EOM-CCSD strategy to compute singlet (S) and triplet (T) energies. 
($-2$) and ($0$) indicate the molecular charge, while a,b,c,$\ldots$ and i,j,k,$\ldots$ denote the virtual and occupied orbital labels, respectively.}
\label{fig:scheme_ccsd}
\end{figure}

In order to investigate the open-shell electronic structures of nanographene systems, we utilize the double ionization potential (DIP) formalism,
\begin{equation}
   \hat{R}^{DIP} =\frac 12 \sum_{ij} r_{ij}  \hat a_j  \hat a_i + \frac 16 \sum_{ijka} r_{ijk}^a  \hat a^{\dagger}_a  \hat a_k  \hat a_j  \hat a_i + ... = \hat{R}_{2h} + \hat{R}_{3h1p} ..., 
\end{equation}
where we restricted ourselves to 2 holes (2h) and 3 holes and 1 particle (3h1p). We consider organic compounds with negligible spin-orbit coupling interactions.
Thus, we can independently optimize the cases for the following spin projection manifolds: $S_z=0$, $S_z= -1$, and $S_z = -2$, and we focus on $S_z = 0$ to target singlet and triplet states within one single DIP-EOM-CCSD calculation.
These states are accessed by performing the double ionization from the dianionic reference HF wavefunction (charge $-2$), a strategy that proved highly reliable.~\cite{dip-eomfpccsd-jctc-2025}
A schematic overview of the approach is shown in Fig.~\ref{fig:scheme_ccsd}.

\section{Computational details}\label{sec:comput-det}
\subsection{Structure optimization}
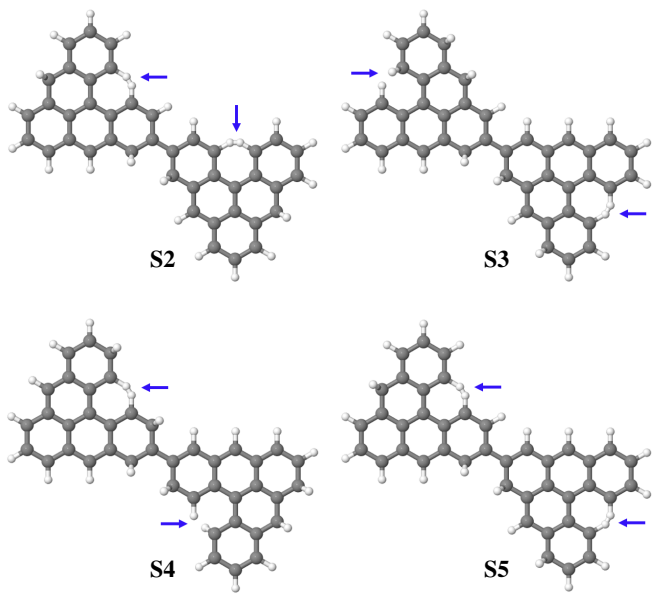
\begin{figure}[h]
\centering

\newcommand{\imglabel}[2]{%
\begin{tikzpicture}
\node[anchor=south west,inner sep=0] (img) at (0,0)
{\includegraphics[width=0.48\columnwidth]{#1}};
\node[anchor=south,font=\bfseries,yshift=3pt]
at (img.south) {#2};
\end{tikzpicture}
}

\imglabel{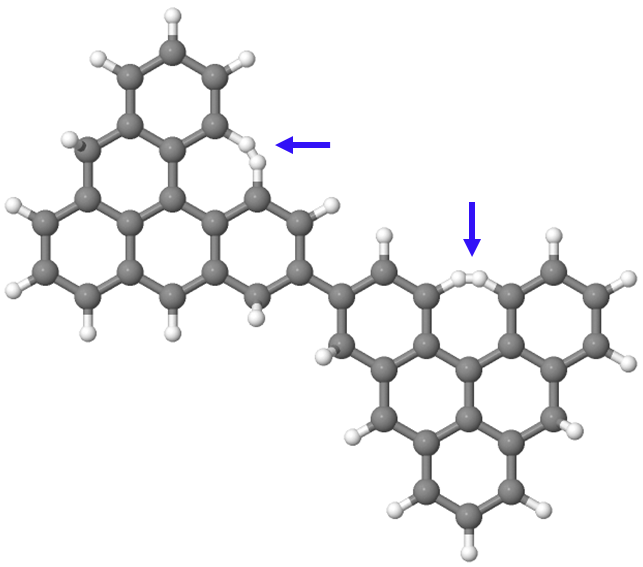}{S2}
\hfill
\imglabel{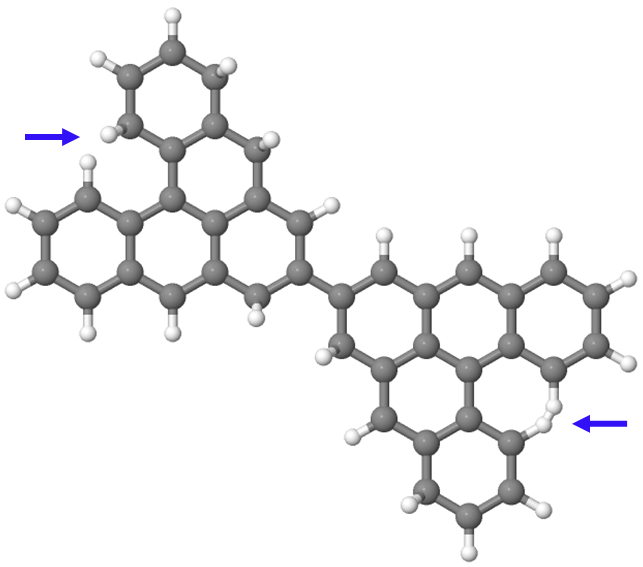}{S3}

\vspace{0.4cm}

\imglabel{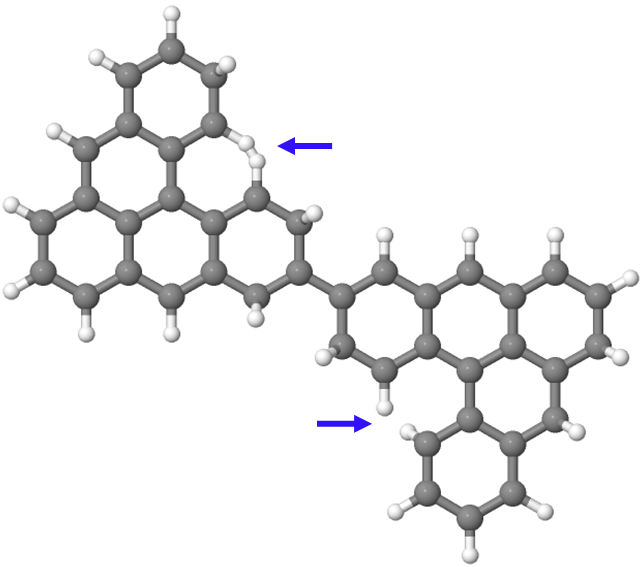}{S4}
\hfill
\imglabel{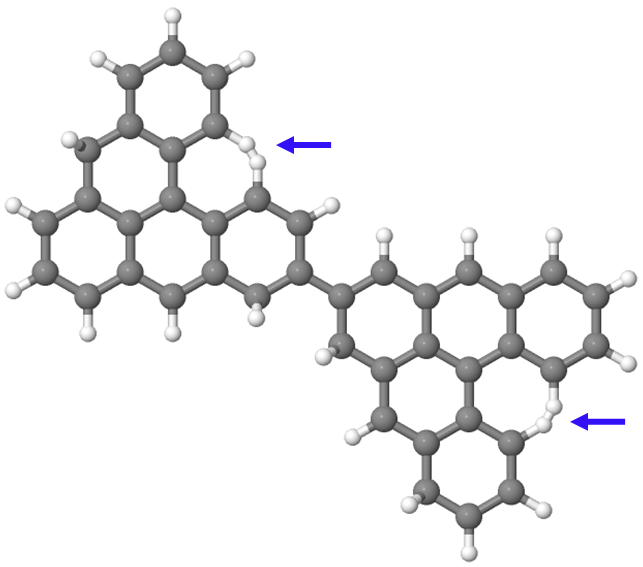}{S5}

\caption{Triangulene dimers with missing carbon atom indicated by blue arrows. At low energies structures can be described by two interacting spin $\frac 12$'s. All structures are based on the \ce{C_42H_24} unit. }
\label{fig:structures_carbon_removal}
\end{figure}

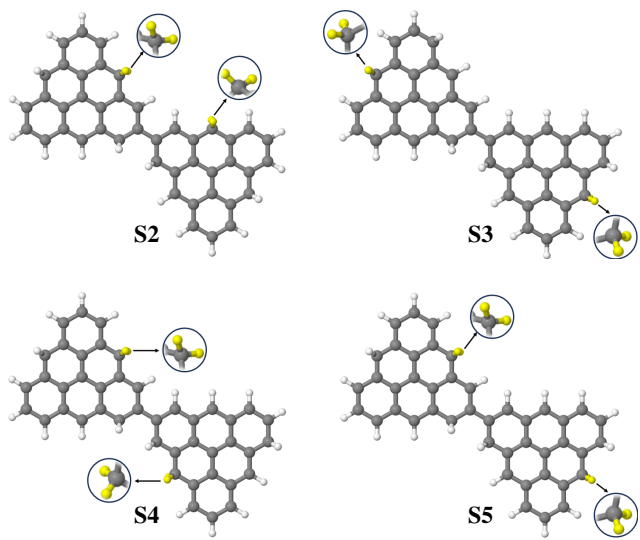
\begin{figure}[h]
\centering

\newcommand{\imglabel}[2]{%
\begin{tikzpicture}
\node[anchor=south west,inner sep=0] (img) at (0,0)
{\includegraphics[width=0.48\columnwidth]{#1}};
\node[anchor=south,font=\bfseries,yshift=3pt]
at (img.south) {#2};
\end{tikzpicture}
}

\imglabel{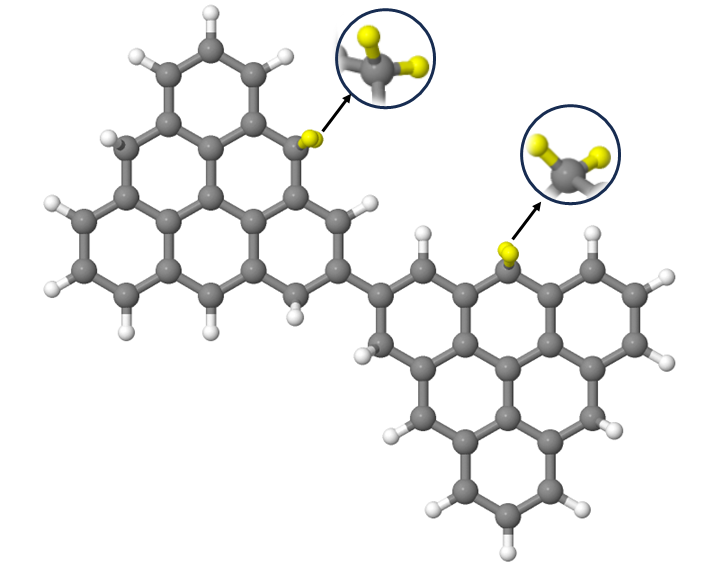}{S2}
\hfill
\imglabel{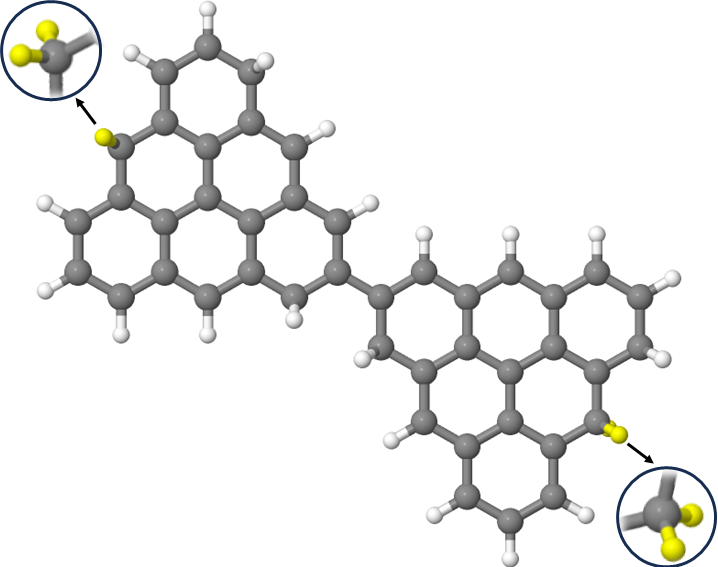}{S3}

\vspace{0.4cm}

\imglabel{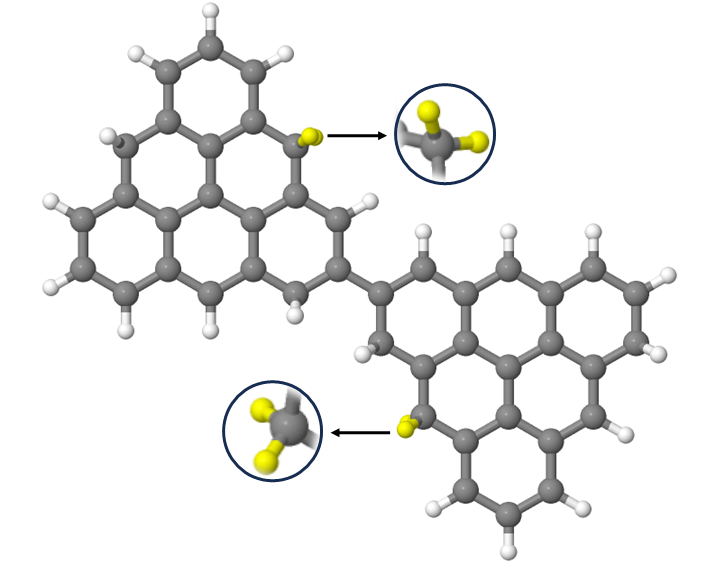}{S4}
\hfill
\imglabel{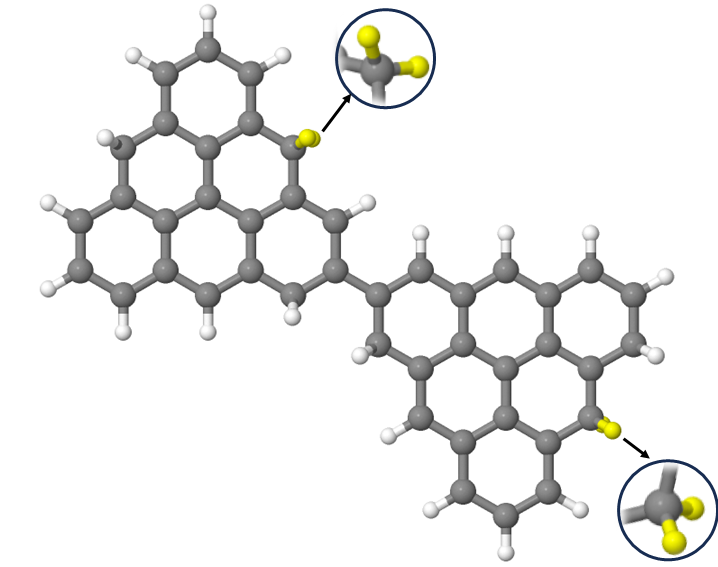}{S5}

\caption{Triangulene dimers with extra hydrogen atom indicated by yellow colors (all structures: \ce{C_44H_24}). At low energies structures can be described by two interacting spin $\frac 12$'s.}
\label{fig:structures_extra_hydrogen}
\end{figure}

We investigate [3]triangulene dimers in which a single carbon atom in each triangulene unit has been dehydrogenated.
The resulting undercoordinated carbon atom binds to a Au atom on the substrate, inducing a bending of the molecule toward the surface.
To model this effect, we considered two approaches: (i) complete removal of the target carbon atom (Fig.~\ref{fig:structures_carbon_removal}) and (ii) saturation of the dehydrogenated site with an additional hydrogen atom (Fig.~\ref{fig:structures_extra_hydrogen}). All molecular visualization in this article was performed using the Jmol v16.3.49 software package ~\cite{jmol-hanson}.

For each case, we examined both fully relaxed and unrelaxed structures. Unrelaxed structures correspond to initial non-optimized geometries constructed in Python v3.12.3~\cite{python-cpython} by arranging all carbon atoms in uniform C-C distances 1.42 \AA. Then, one hydrogen atom was added to each carbon atom at the edge by using Avogadro v.1.90.0 software package~\cite{avogadro-hanwell}. Relaxed structures are obtained by optimization of the geometry of unrelaxed structures. These structures were optimized at the density functional theory (DFT) level using the ORCA v6.0.1 software package~\cite{orca-2012,orca-2026} with the BP86 exchange-correlation functional~\cite{bp86-becke,bp86-perdew} and the def2-SVP basis set for all atoms.
Upon relaxation, the overall structural features of the dimers were well preserved, with only minor geometric adjustments observed around the carbon-deficient region.
The corresponding xyz coordinates are provided in Supplementary Material.
\subsection{DIP-EOM-CCSD calculations}
All DIP-EOM-CCSD calculations were performed using the open-source PyBEST v2.1.0 software package.~\cite{boguslawski2021pythonic, boguslawski2024pybest, pybest-gpu-jctc-2024}.
In all calculations, the 1s core orbitals of carbon were frozen, and the two-electron integrals were approximated using Cholesky decomposition~\cite{cholesky-review-2011} with a threshold of $10^{-5}$, which is more than sufficient for accurate excitation and relative energies.
All structures were investigated employing the STO-3G and 6-31G* basis sets. For the olympicene dimers, additional calculations were performed using the cc-pVDZ basis set.

\section{Olympicene dimer}
\begin{figure}[h]
\centering
\newcommand{\cornerlabel}[2]{%
\begin{tikzpicture}
\node[anchor=south west, inner sep=0] at (0,0)
{\includegraphics[width=0.7\columnwidth]{#1}};
\node[anchor=south east, font=\bfseries, xshift=-3pt, yshift=3pt]
at (current bounding box.south east) {#2};
\end{tikzpicture}
}
\cornerlabel{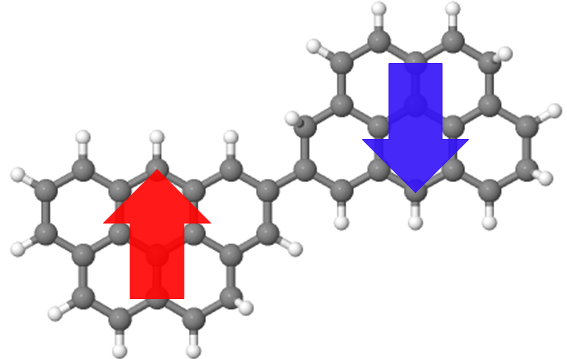}{S1}
\caption{Dimer olympicenes: \ce{C_38H_20} (S1). Low energy properties can be described by two spin-$\frac 12$ antiferomagnetic Heisenberg model with exchange coupling $J$. }
\label{fig:structures_representative}
\end{figure}
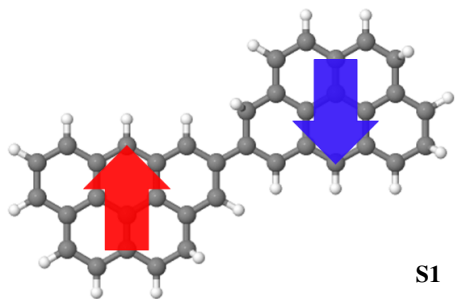

\begin{figure}[h]
\centering

\begin{minipage}{0.49\columnwidth}
\centering
\textbf{HOMO}
\end{minipage}
\hfill
\begin{minipage}{0.49\columnwidth}
\centering
\textbf{HOMO-1}
\end{minipage}

\vspace{0.2cm}

\begin{tikzpicture}

\node[inner sep=0] (L)
{\includegraphics[width=0.45\columnwidth]{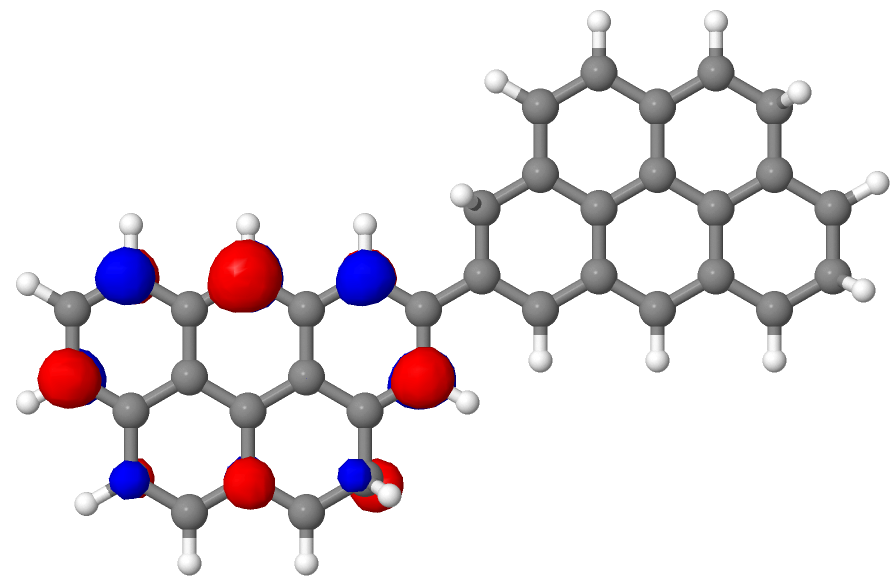}};

\node[inner sep=0] (R) at (0.5\columnwidth,0)
{\includegraphics[width=0.45\columnwidth]{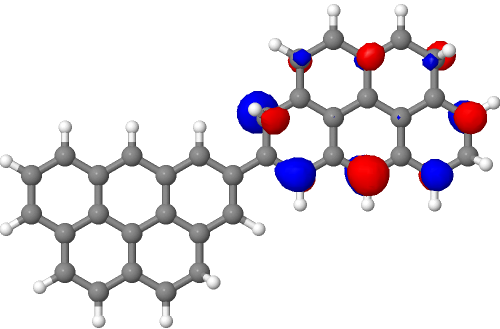}};

\node[anchor=south,font=\bfseries,xshift=-6pt,yshift=3pt]
at (R.south east) {S1};

\end{tikzpicture}

\caption{Comparison of the HOMO and HOMO-1 orbitals (form the HF calculations and charge $-2$) in the unrelaxed structures of \ce{C_38H_20} (S1) (nos. 125, 124, cutoff 0.05).}
\label{fig:HF_orbitals_dimer_olympicene}
\end{figure}

We start our analysis of the method with Dimer olympicenes shown in Fig. \ref{fig:structures_representative}.
These are molecular structures formed by linking two units of olympicene, a polycyclic aromatic hydrocarbon composed of five fused benzene rings arranged in a pattern reminiscent of the Olympic rings. 
A key feature of olympicene is its open-shell character, meaning that its ground state can host one unpaired electron per molecule due to its particular $\pi$-electron topology. 
When two such units form a dimer, each contributes an unpaired electron, a free spin. 
The corresponding HOMO and HOMO-1 orbitals (form our HF calculations) are shown in Fig.~\ref{fig:HF_orbitals_dimer_olympicene}.
These orbitals are localized on edges, and mainly on one of the two sublattices from a honeycomb lattice. The spins filling these states interact through antiferromagnetic exchange coupling, leading to an overall singlet ground state and a triplet as the first excited state.
A singlet-triplet splitting is equal to a spin exchange $J$ interaction in an effective two-site spin-$\frac 12$ Heisenberg Hamiltonian, shown explicitly in Appendix A. 
Table \ref{tab:dimer_olympicene} shows an effective $J$ interaction of relaxed and unrelaxed structures, and for different basis choices.
Relaxed structures reduce $J$ by more than 25$\%$ but we expect that in realistic conditions the underlying substrate stabilizes the structure and the relaxation of the structures is not so strong. 
The least computationally expensive basis is STO-3G, a minimal basis set (a linear combination of 3 primitive Gaussian functions), but at the same time it is believed to be least accurate. 
The most expensive basis is cc-pVDZ, which is comparable in accuracy with the 6-31G* basis set. The obtained values of unrelaxed structures are smaller by more than $30\%$ compared to the experimental value from Ref. \onlinecite{zhao2025spin}, $J^{Oli}_{exp}=38$ meV. We attribute this discrepancy to the lack of screening in our calculations. Screening from the Au substrate is expected to reduce the effect of long-range Coulomb interactions and enhance the relative importance of the short-range Hubbard interaction. 

\begin{table}[!h]
\begin{center}
\begin{tabular}{|p{2.8cm}||p{2.5cm}|p{2.5cm}|}
\hline
\begin{tikzpicture}[baseline=(current bounding box.center)]
\coordinate (a) at (0,1cm);
\coordinate (b) at (2.8cm,0);
\draw (a) -- (b);

\node[anchor=north west, inner sep=0.5pt] at ([xshift=28pt,yshift=-2pt]a) {Structure (S1)};
\node[anchor=south east, inner sep=0.5pt] at ([xshift=-55pt,yshift=3pt]b) {Basis};
\end{tikzpicture}
& \centering Unrelaxed geometry (meV) & \centering Relaxed geometry (meV)

\tabularnewline
\hline\hline
\centering STO-3G & \centering 41,0 & \centering 24,8 \tabularnewline
\centering 6-31G* & \centering 25,3 & \centering 18,0 \tabularnewline
\centering cc-pVDZ & \centering 27,0 & \centering 19,4 \tabularnewline

\hline
\end{tabular}
\end{center}
\caption{DIP-EOM-CCSD singlet-triplet splitting in relaxed and unrelaxed structures of olympicene dimer \ce{C_38H_20}.}\label{tab:dimer_olympicene}
\end{table}

\section{Triangulene dimers}
Despite the discrepancy between our results and the experimental exchange coupling J for the olympicene dimer, our approach still allows us to compare $J$ in different triangulene dimer structures with dehydrogenated target sites, taking into account that the obtained values are underestimated. We consider four structures with different positions of removed carbon/extra hydorgeneted of target sites, shown in Fig. \ref{fig:structures_carbon_removal} and Fig. \ref{fig:structures_extra_hydrogen}, respectively, and analyze how dehydrogenation affects exchange splitting. In particular, structure $S2$ has two dehydrogenated sites on the same side and the closest to each other. On the other hand, structure $S3$ has two dehydrogenated sites on opposite sides of the structure. Structure $S4$ and $S5$ are somewhere in-between. A dehydrogentation position affects HOMO and HOMO-1 charge distribution. This is shown in Fig. \ref{fig:homo_remove_carbon} for carbon removal structures and in Fig. \ref{fig:homo_add_carbon} for structures with extra hydrogens on target sites. Clearly, an electronic density avoids target sites and in consequence HOMO and HOMO-1 charges are closest to each other for $S3$ structure and farthest from each other for $S2$. Calculated exchange interaction are presented in Table \ref{tab:all_values} for relaxed and unrelaxed structures, and STO-3G and 6-31G* basis sets.
Similarly to the results for olympicine dimer, carbon-removal and hydrogen-passivated strutures give similar results. The relaxation of structures reduces significantly $J$ in all cases. 

We analyze now 6-31G* basis results as this basis is expected to be more reliable than STO-3G. For structure S2 and S4 we obtain $J$ below 3.0 meV for relaxed structures. In contrast, for the same basis, the relaxed structure S3 gives $J$ above 40 meV, while if the structure is unrelaxed, $J$ is above 90 meV for hydrogen-passivated sites. This reflects the fact that HOMO and HOMO-1 charge densities practically fully overlap in this case (hydrogen-passivated sites), see Fig. \ref{fig:homo_add_carbon}, while the charge densities for S3 are slightly different for the structure with carbon removal, Fig. \ref{fig:homo_remove_carbon}.
The charge density of HOMO and HOMO-1 for S4 is far from inter-triangulene connection what translates to quite small $J$, and comparable to the values for S2.
Structure S5 has the charge density of HOMO far from inter-triangulene connection, while close to it for HOMO-1, thus $J$ is between the value for S2 or S4, and the value for S3, which is the largest among all considered configurations.

\newcommand{\pairrow}[3]{%
\begin{tikzpicture}

\node[inner sep=0] (L) at (0,0)
{\includegraphics[width=0.4\columnwidth]{#1}};

\node[inner sep=0] (R) at (0.5\columnwidth,0)
{\includegraphics[width=0.4\columnwidth]{#2}};

\node[anchor=south east, font=\bfseries,
xshift=-1pt, yshift=-1pt]
at (R.south east) {#3};

\end{tikzpicture}

\vspace{0.25cm}
}

\begin{figure}[h]
\centering

\begin{minipage}{0.49\columnwidth}
\centering
\textbf{HOMO}
\end{minipage}
\begin{minipage}{0.49\columnwidth}
\centering
\textbf{HOMO-1}
\end{minipage}

\vspace{0.3cm}

\pairrow{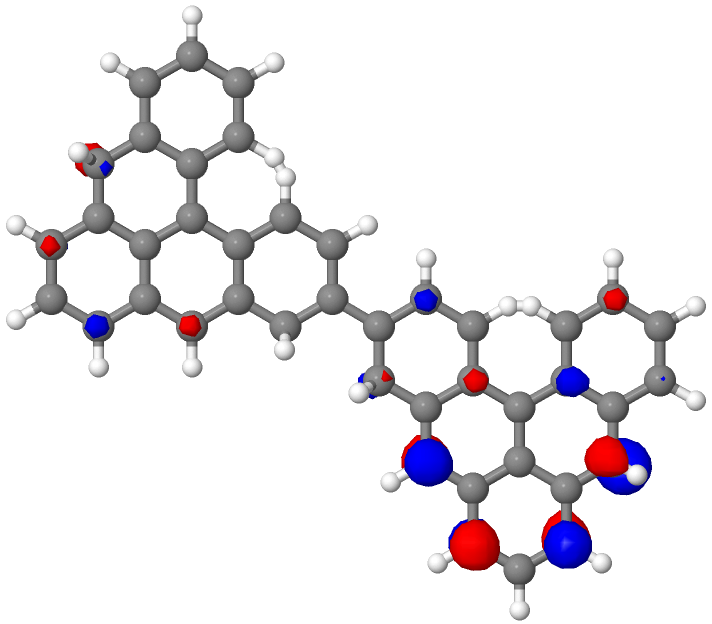}
        {c42h24v2_mo138.png}
        {S2}

\pairrow{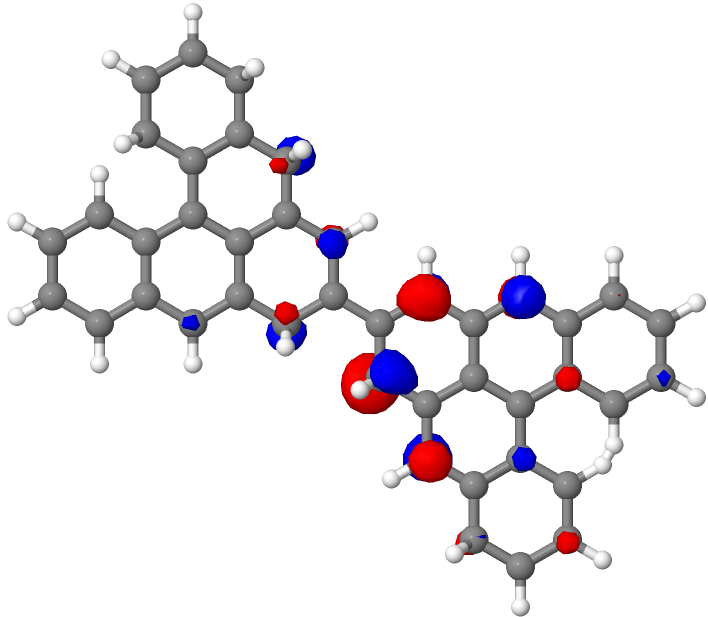}
        {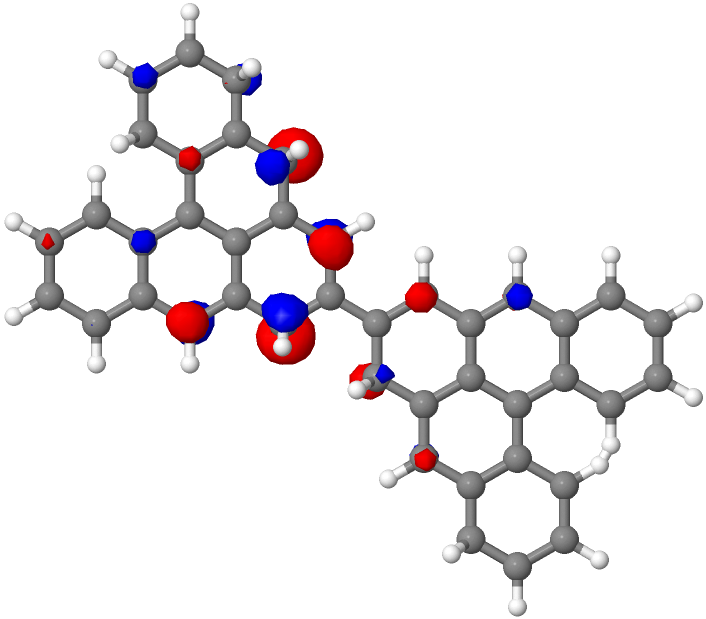}
        {S3}

\pairrow{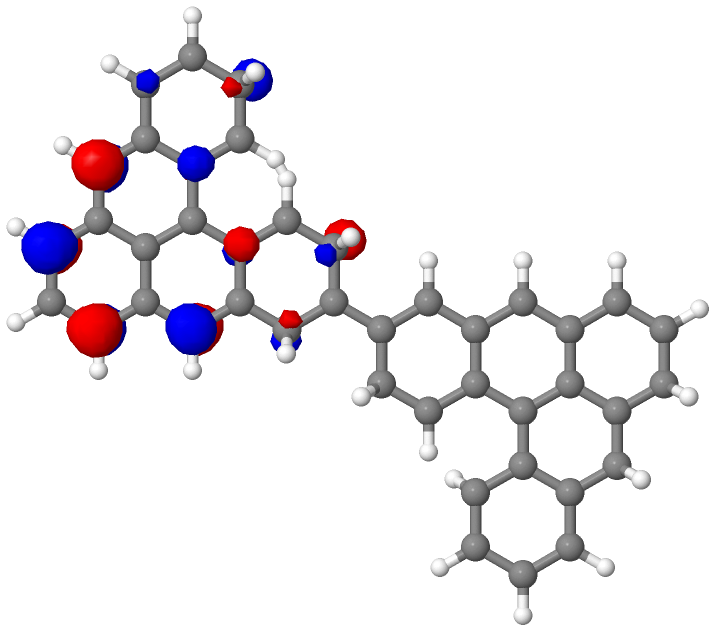}
        {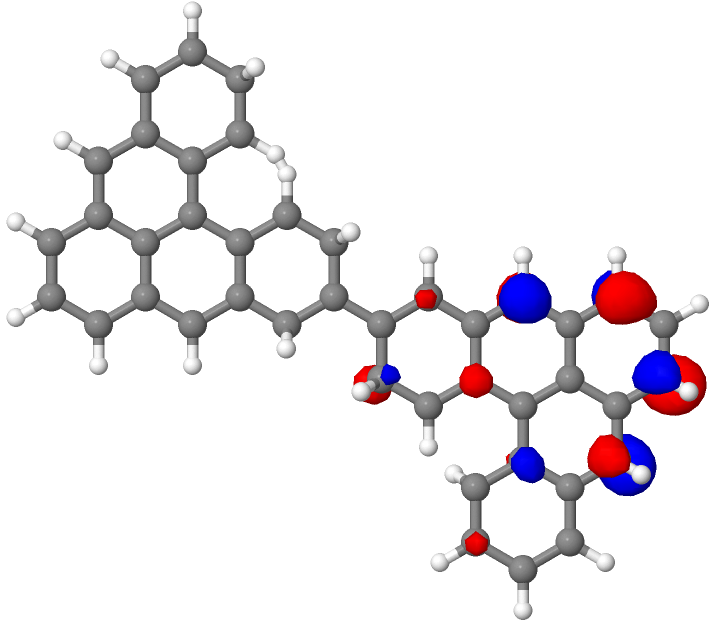}
        {S4}

\pairrow{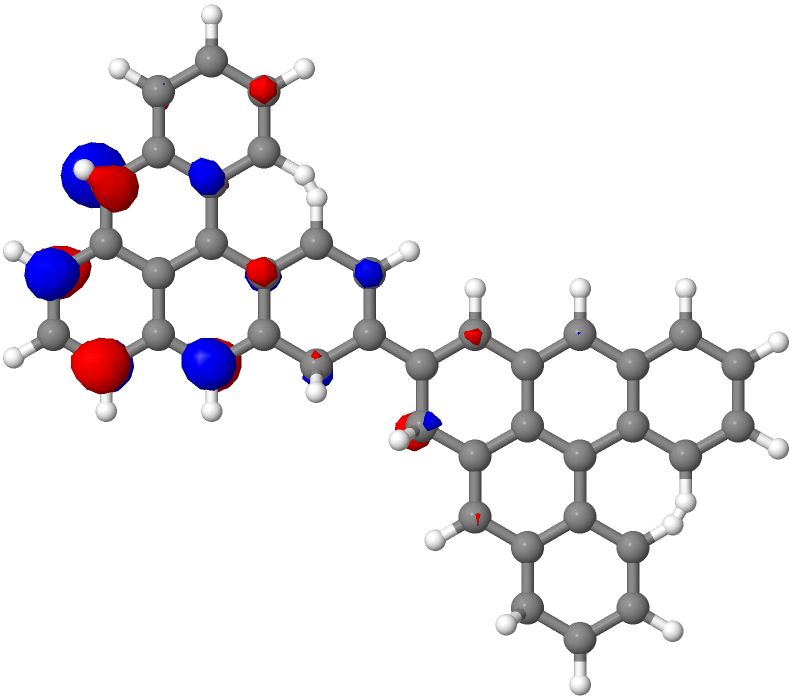}
        {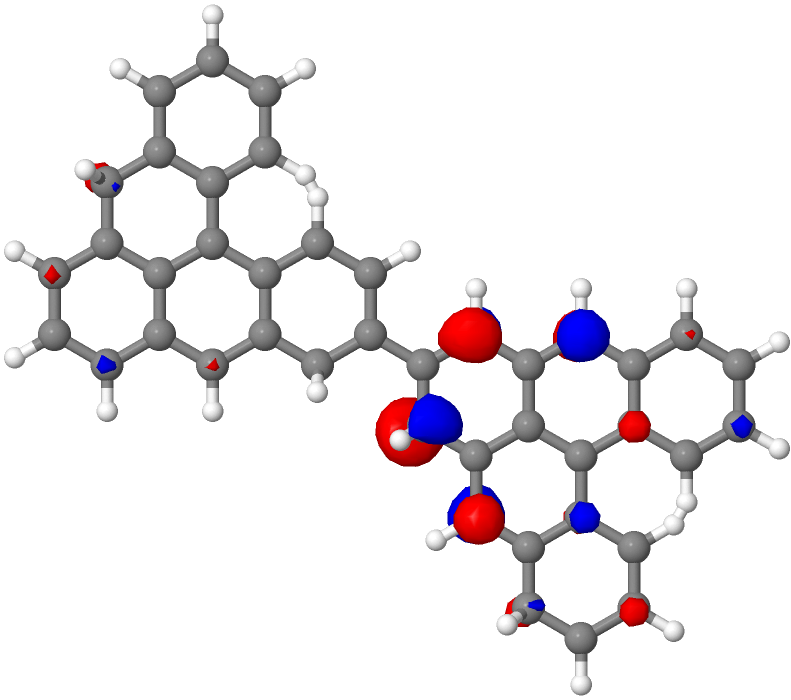}
        {S5}

\caption{HOMO and HOMO-1 orbitals (form the HF calculations and charge $-2$) comparison in unrelaxed structures of selected carbon removed sites \ce{C_42H_24} (nos. 139, 138, cutoff 0.05)}
\label{fig:homo_remove_carbon}
\end{figure}

\renewcommand{\pairrow}[3]{%
\begin{tikzpicture}

\node[inner sep=0] (L) at (0,0)
{\includegraphics[width=0.4\columnwidth]{#1}};

\node[inner sep=0] (R) at ({0.5\columnwidth},0)
{\includegraphics[width=0.4\columnwidth]{#2}};

\node[anchor=south east, font=\bfseries,
xshift=-1pt, yshift=-1pt]
at (R.south east) {#3};

\end{tikzpicture}
\vspace{0.25cm}
}

\begin{figure}[h]
\centering

\begin{minipage}{0.49\columnwidth}
\centering
\textbf{HOMO}
\end{minipage}
\begin{minipage}{0.49\columnwidth}
\centering
\textbf{HOMO-1}
\end{minipage}

\vspace{0.3cm}

\pairrow{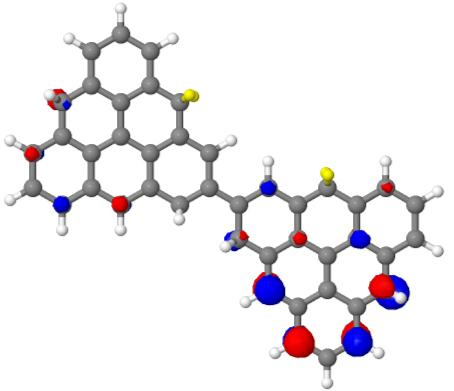}
        {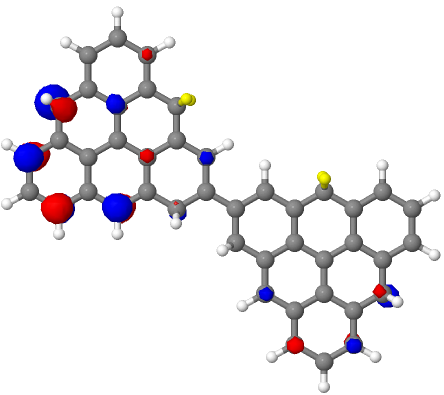}
        {S2}

\pairrow{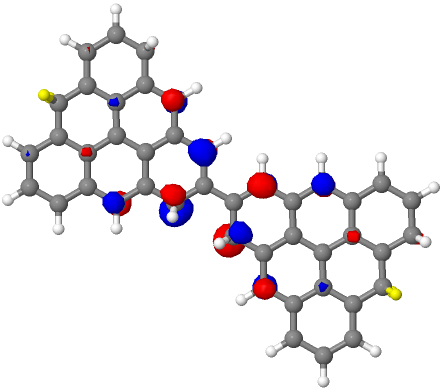}
        {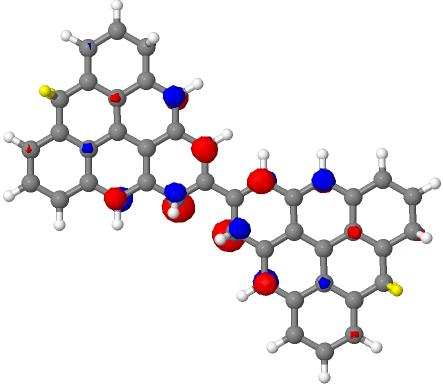}
        {S3}

\pairrow{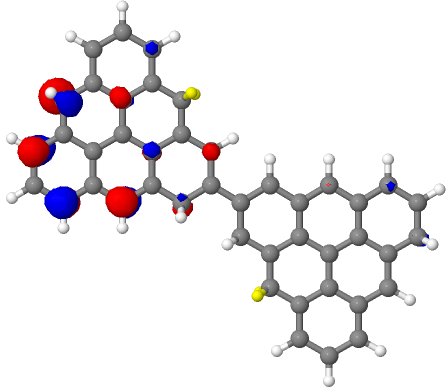}
        {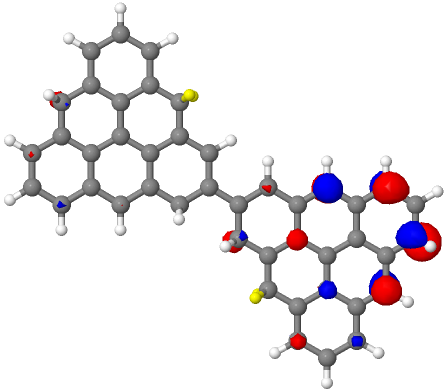}
        {S4}

\pairrow{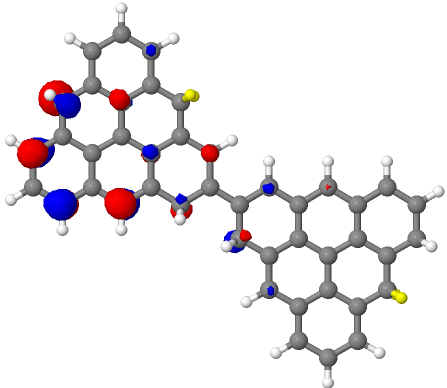}
        {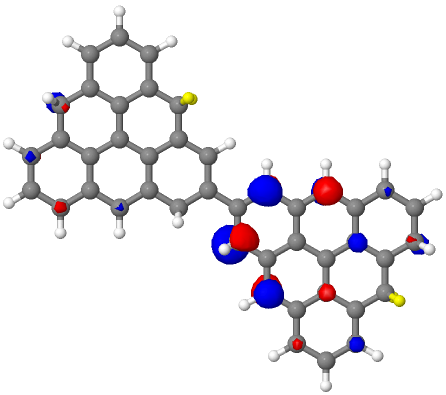}
        {S5}

\caption{HOMO and HOMO-1 orbitals (form the HF calculations and charge $-2$) comparison in unrelaxed structures of hydrogen-passivated \ce{C_44H_24} (nos. 145, 144, cutoff 0.05)}
\label{fig:homo_add_carbon}
\end{figure}

\begin{table}
\begin{center}
\begin{tabular}{|p{2.4cm}|p{1.22cm}p{1.22cm}p{1.22cm}p{1.22cm}|}
\hline
\begin{tikzpicture}[baseline=(current bounding box.center)]
\coordinate (a) at (0,1.0cm);
\coordinate (b) at (2.4cm,0);
\draw (a) -- (b);

\node[anchor=north west, inner sep=0.5pt] at ([xshift=30pt,yshift=-4pt]a) {Structures};
\node[anchor=south east, inner sep=0.5pt] at ([xshift=-40pt,yshift=5pt]b) {Basis};
\end{tikzpicture}
& \centering S2 (meV) & \centering S3 (meV) & \centering S4 (meV) &  \centering S5 (meV) \tabularnewline
\hline\hline
\multicolumn{5}{|c|}{\textbf{Geometry with carbon removal (\ce{C_42H_24})}} \\
\hline
\centering STO-3G &
\begin{tabular}{c}10,7\\ \hline \opt{6,7}\end{tabular} &
\begin{tabular}{c}78,2\\ \hline \opt{49,8}\end{tabular} &
\begin{tabular}{c}14,0\\ \hline \opt{6,9}\end{tabular} &
\begin{tabular}{c}27,7\\ \hline \opt{18,2}\end{tabular} \\
\hline

\centering 6-31G* &
\begin{tabular}{c}5,2\\ \hline \opt{2,6}\end{tabular} &
\begin{tabular}{c}78,8\\ \hline \opt{43,8}\end{tabular} &
\begin{tabular}{c}6,2\\ \hline \opt{2,9}\end{tabular} &
\begin{tabular}{c}19,1\\ \hline \opt{10,2}\end{tabular} \\
\hline\hline
\multicolumn{5}{|c|}{\textbf{Geometry with hydrogen-passivated (\ce{C_44H_24})}} \\
\hline
\centering STO-3G &
\begin{tabular}{c}10,3\\ \hline \opt{6,0}\end{tabular} &
\begin{tabular}{c}88,3\\ \hline \opt{50,7}\end{tabular} &
\begin{tabular}{c}10,0\\ \hline \opt{6,2}\end{tabular} &
\begin{tabular}{c}27,6\\ \hline \opt{16,9}\end{tabular} \\
\hline
\centering 6-31G* &
\begin{tabular}{c}4,8*\\ \hline \opt{ }\end{tabular} &
\begin{tabular}{c}92,5\\ \hline \opt{46,3}\end{tabular} &
\begin{tabular}{c}5,0 \\ \hline \opt{2,4}\end{tabular} &
\begin{tabular}{c} \\ \hline \opt{8,9}\end{tabular} \\
\hline
\end{tabular}
\vspace{2mm}
\begin{tabular}{ll}
\fcolorbox{black}{white}{\rule{0pt}{1.4ex}\rule{1.4ex}{0pt}}
& Unrelaxed structures\\

\fcolorbox{black}{gray!20}{\rule{0pt}{1.4ex}\rule{1.4ex}{0pt}}
& Relaxed structures\\
\end{tabular}

\end{center}
\caption{Singlet-triplet splitting in relaxed and unrelaxed structures of selected carbon removed sites and hydrogen-passivated.}\label{tab:all_values}
\end{table}

\section{Conclusions}
We have calculated singlet-triplet gaps for many tip-induced dehydrogentated [3]-triangulene dimers using all electron DIP-EOM-CCSD method.
A dehydrogeneted carbon $p_z$ orbital creates a bond with underlying substrates atom thus does not contribute to low energy properties of the structure. In our models, we implement this effect in two ways: by removing a dehydrogenated carbon atom from the structure or additionally passivating it by hydrogen (cf. structures~\ref{fig:structures_carbon_removal} and~\ref{fig:structures_extra_hydrogen}). 
For all considered structures with different positions of removed/extra hydrogen passivated sites, both models give comparable results. A significant difference is noticed between unrelaxed and relaxed structures. In experiments, triangulene dimers are created using on-surface synthesis, so underlying substrate is expected to reduce the effect of geometry relaxation. We show that the position of the removed/hydrogen passivated site strongly affects the singlet-triplet splitting, an effective spin exchange coupling $J$. If target sites are farthest from each other like in structure $S3$, they shift the density of unpaired spins closer to the inter-triangulene connection significantly increasing $J$ giving the largest values. On the opposite limit is structure $S2$, where target sites are closest to each other and spin densities are shifted apart from each other. Interestingly, in these two limited situations, $J$ differ by an order of magnitude, from few meV to tens of meV; our calculations do not include the effect of the substrate, which would affect the structure relaxation effect and the Coulomb interaction. Screening would reduce the interaction to make it short-range, and as was shown in a previous work \cite{saleem2024superexchange}, this would increase exchange interaction. Nevertheless, we have shown that the dehydrogenation seems to be a promising path toward designing spin models with a desired interaction strength. Additionaly, the dehydrogentation could also be used to create alternating spin models, structures with spins that vary from site to site, or even hybrid spin models, with spins of different magnitude, according to a recent proposition by one of us presented in Ref. \onlinecite{saleem2025engineering}.

\begin{acknowledgments}
Our calculations were performed at the Wrocław Center for Networking and Supercomputing, grant no. 317. 
P.T.~acknowledge financial support from the SONATA BIS research grant from the National Science Centre, Poland (Grant No. 2021/42/E/ST4/00302).
\end{acknowledgments}

\bibliography{main.bib}

\appendix

\section*{Appendix - spin-$\frac 12$ dimer}
Structures considered in the main article, at low energies can be described by a two site spin-1/2 Hamiltonian given by 
\begin{equation}
    H_{s=\frac{1}{2}} = J{\textbf{s}}_1\cdot{\textbf{s}}_2,
\end{equation}
where $\textbf{s}_{1,2}$ are spin operators and
written in terms of raising and lowering operators has a form
\begin{equation}
    H_{s=\frac{1}{2}} = J\left[\frac{1}{2}\left(\hat{s}^+_1s^-_2+\hat{s}^-_1\hat{s}_2^+\right)+\hat{s}_1^z\hat{s}_2^z\right].
\end{equation}
Each spin-1/2 site has two possible states, $\ket{\uparrow}$ and $\ket{\downarrow}$. For the two sites, there are a total of four states, which reduce to two when restricting to the $(s_1^z+s_2^z)=0$ subspace. This yields a 2x2 Hamiltonian
\begin{equation}
     \textbf{H}_{s=\frac{1}{2}} = \frac{J}{4} \begin{pmatrix}
-1 & 2 \\
2 & -1
\end{pmatrix}.
\label{eq:2sitespin12}
\end{equation}
Diagonalizing this Hamiltonian one finds a singlet ground state with energy $E_S=-\frac 34 J$, and a triplet excited state $E_T=\frac 14 J$ with a singlet-triplet $J=E_T-E_S$. 

\begin{widetext}

\section*{Supplementary Material.\\ Cartesian coordinates of all structures in Angstroms (\AA)}

\verbatiminput{All_structures.txt}

\end{widetext}

\end{document}